\documentclass[lettersize,journal]{IEEEtran}
\usepackage{amsmath,amsfonts}
\usepackage{algorithmic}
\usepackage{algorithm}
\usepackage{svg}
\usepackage{array}
\usepackage[caption=false,font=normalsize,labelfont=sf,textfont=sf]{subfig}
\usepackage{textcomp}
\usepackage{stfloats}
\usepackage{url}
\usepackage{graphicx}
\usepackage{tabularx}
\usepackage{amsmath,amssymb}
\usepackage{verbatim}
\usepackage{graphicx}
\usepackage{cite}
\newtheorem{assumption}{Assumption}
\newtheorem{remark}{Remark}

\usepackage[most]{tcolorbox}

\newtcbtheorem{them}{\textbf{Them}}{
	enhanced,
	breakable,
	colback=white,
	colframe=black,
	boxrule=0.8pt,
	arc=1.5mm,
	fonttitle=\bfseries,
	coltitle=black,
	attach boxed title to top left={
		xshift=3mm,
		yshift=-2mm
	},
	boxed title style={
		colback=white,
		colframe=white
	},
	left=3mm,
	right=3mm,
	top=3mm,
	bottom=3mm
}{them}

\begin{document}
	
	\title{Safeguarding Intelligent Power Dispatch: \\A Performance-Equivalent-Model-Based Safety-Critical Control Method for DC Microgrids }
	
\author{
	Shaoliang ZHOU,~\IEEEmembership{Student Member,~IEEE},
	and Ziqing ZHU$^{*}$,~\IEEEmembership{Member,~IEEE}
	\thanks{$^{*}$Corresponding author: Ziqing ZHU (e-mail: ziqzhu@polyu.edu.hk).}
}
	\markboth{Journal of \LaTeX\ Class Files,~Vol.~14, No.~8, August~2021}%
	{Shell \MakeLowercase{\textit{et al.}}: A Sample Article Using IEEEtran.cls for IEEE Journals}

	
	\maketitle
	
	\begin{abstract}
    Intelligent power dispatch is enabling fast and self-adaptive decision-making in microgrids, yet its deployment in real-world industry remains limited due to unresolved safety concerns. Existing safety-preserving approaches typically rely on an accurate characterization of the feasible operating region, which is impossible to obtain under uncertain and heterogeneous system dynamics. Moreover, even a dispatch command verified as feasible at a discrete setpoint may become unsafe during its continuous physical execution. To address these challenges, this paper proposes a safety-aware dispatch adjustment framework that uses performance-equivalent modeling of closed-loop systems (PEM-CLS) and embeds high-order control barrier functions and control Lyapunov functions within a quadratic programming formulation (HOCBF-CLF-QP). PEM-CLS captures EMS-required power-regulation dynamics solely from port-level responses, without requiring internal subsystem models, parameters, or controller information. Its modeling error is analytically characterized and explicitly bounded, enabling requirement-oriented dynamic abstraction for safety-critical control. Based on these PEMs, the HOCBF-CLF-QP safety controller coordinates heterogeneous transient power responses and corrects unsafe EMS commands to maintain dynamic safety during command execution. Four deliberately constructed extreme stress-test scenarios are used to expose conflicts between nominal EMS commands and physical safety. Without intervention, the DC-bus voltage violates its safety limit in all four cases, whereas the proposed framework maintains QP feasibility and limits the voltage deviation. The results demonstrate that the proposed framework provides an algorithm-agnostic safety execution layer for intelligent power dispatch without requiring an exact feasible-region model or detailed internal information of lower-level devices.
	\end{abstract}
	
	
	\section{Introduction}
    \IEEEPARstart{A}{s} the penetration of renewable energy continues to increase in microgrids, and system operation is becoming increasingly uncertain, making conventional energy management based on fixed rules or operator experience progressively inadequate. Intelligent power dispatch has therefore emerged as an important direction for microgrid energy management systems (EMSs). By adapting power commands to changing operating conditions, intelligent power dispatch can substantially improve decision-making under complex system conditions. Existing research has evolved from explicit-model-based optimization toward learning-based decision-making, offering increasingly powerful capabilities and new possibilities for more autonomous microgrid operation.
    
    Despite these advances, the deployment of intelligent power dispatch in real-world industrial systems remains highly cautious. A fundamental obstacle is that existing approaches to safety-preserving intelligent power dispatch typically rely on an accurate characterization of the feasible operating region and constrain power dispatch decisions accordingly. In practical microgrids, however, this region varies with operating conditions, while its boundary depends on system dynamics that cannot be fully identified or modeled with sufficient accuracy. Consequently, satisfying the safety constraints represented in the power dispatch model does not necessarily guarantee that the resulting decision remains safe in the physical system \cite{Z1lang2026safety,Z2guo2024multi}. This difficulty is further compounded by the fact that the physical responses of heterogeneous microgrid devices cannot be completely represented by steady-state operating constraints \cite{Z2guo2024multi}. More importantly, even if a dispatch command can be accurately verified as feasible at a given operating point, its physical execution is not necessarily safe. EMS commands are updated at discrete instants, whereas physical devices respond continuously according to their own dynamics. Differences in response bandwidths and ramping capabilities may therefore create transient power imbalances during command transitions, leading to DC-bus voltage deviations or even violations of the prescribed safety limits \cite{Z3zhang2021voltage}. Meanwhile, renewable generation and load demand continue to vary between consecutive dispatch instants, causing the actual system state to deviate from the operating condition anticipated by the EMS \cite{Z4macavs2020impact}. As a result, a dispatch decision that is feasible at the  power dispatch instant may still drive the system outside its safe operating region during continuous physical execution \cite{Z2guo2024multi}.

    To compensate for the above gap in continuous dynamic safety, DC microgrids commonly rely on additional safety-support resources, fast control strategies, and conservative dispatch mechanisms. However, owing to the high-dimensional nonlinear characteristics of dynamic security constraints and the associated online computational burden, it is difficult for EMSs to continuously solve dynamic safety boundaries that remain strictly consistent with the real-time operating state within a limited dispatch interval \cite{Z7deng2025rapid,Z8wang2025dynamics}. Consequently, relatively large safety margins are typically imposed in practical operation to compensate possible safety risks caused by unclear modeling dynamics. Such conservative margins prematurely restrict the regulation capability of renewable generation and other flexible resources, wasting part of the renewable power that would remain admissible under the actual dynamic constraints from participating in dispatch and thereby reducing the overall renewable energy efficiency \cite{Z9yi2023scenario}. 

    The key to the above issue does not lie in simply incorporating more dynamic processes into the EMS optimization but rather in developing a trustworthy intermediate model that bridges dispatch decisions of EMSs and the physical dynamic responses of devices in a microgrid system. Such a model should preserve the key multi-timescale dynamic characteristics that govern safety-related behavior while explicitly quantifying the associated model errors so that the actual dynamic responses induced by dispatch commands can be predicted with sufficient reliability. 

    To reliably transfer safety guarantees from the power dispatch model to the physical system, the intermediate model must satisfy two essential requirements. First, it must preserve the dynamics that determine whether safety constraints are violated; otherwise, accurate steady-state prediction alone cannot reveal unsafe behavior during command execution \cite{Z7deng2025rapid,Z11pramanik2025assessing}. Second, its modeling error must be explicitly bounded; otherwise, the safety margin defined in the model cannot be translated into a valid guarantee for the physical system \cite{Z10awan2025reduced}. Therefore, high average prediction accuracy alone is insufficient for safety-critical control, since significant local errors may still occur during critical transients or near constraint boundaries \cite{Z11pramanik2025assessing}. Existing modeling approaches struggle to satisfy both requirements simultaneously. High-fidelity physics-based models can capture detailed system dynamics, but their high dimensionality and computational burden hinder online safety assessment \cite{Z7deng2025rapid}. Reduced-order models and data-driven models improve computational efficiency, but may omit dynamics that are critical to safety, while their approximation errors are difficult to bound reliably \cite{Z10awan2025reduced,Z11pramanik2025assessing}. 

   To overcome these barriers to the practical deployment of intelligent power dispatch, this paper develops an algorithm-agnostic safety-control framework that guarantees dynamic safety during dispatch execution without requiring an exact model of the feasible operating region or access to proprietary information from lower-level devices. The main contributions are summarized as follows.

    \begin{itemize}

    \item \textbf{An algorithm-agnostic safety execution framework for intelligent power dispatch.}
    An independent safety execution layer is introduced between the scheduler and the physical microgrid. The proposed layer depends only on the power commands issued by the upper-level scheduler and the corresponding physical responses, allowing the same safety mechanism to be applied to any power dispatch method that satisfies the required command interface, regardless of its internal decision-making structure. Before a command is physically executed, its dynamic safety is assessed. If the original command may lead to unsafe operation, only the intervention necessary to restore safety is introduced while preserving the original power dispatch decision as closely as possible. The proposed framework therefore relieves intelligent power dispatch algorithms from having to independently certify the safety of the physical system, substantially lowering a major barrier to their deployment in industrial microgrids.

    \item \textbf{Verifiable dynamic modeling without detailed internal information.}
    To avoid relying on an accurately modeled feasible operating region, this paper develops a performance-equivalent modeling method for closed-loop systems. The proposed method requires no knowledge of subsystem topology, controller implementation, or component parameters, but instead extracts from port-level responses only the dynamic behavior required for safety-critical execution. Based on robust control theory and frequency-domain analysis, the existence and non-uniqueness of performance-equivalent models satisfying the prescribed external performance requirements are rigorously established. The results show that each closed-loop subsystem admits a family of valid equivalent models, allowing the model order and fidelity to be selected according to the available computational resources and EMS requirements.

    \item \textbf{Extension of safety guarantees from discrete dispatch points to continuous physical execution.}
    Based on the resulting performance-equivalent models, an HOCBF-CLF-QP safety-control mechanism is developed to constrain the continuous system trajectory during the execution of intelligent dispatch commands. The HOCBF establishes real-time dynamic safety constraints, while the CLF promotes recovery toward the desired operating condition through relaxation variables. These requirements are coordinated through an efficiently solvable quadratic program that modifies the original dispatch command only when necessary. Simulations under severe renewable-power fluctuations demonstrate that the proposed controller maintains the DC-bus voltage within its prescribed safety limits throughout the execution process while preventing the supercapacitor from overcharging or overdischarging.

    \end{itemize}

    The remainder of this paper is organized as follows. Section II presents the proposed performance-equivalent modeling method for closed-loop systems. Section III develops the PEM-based HOCBF--CLF--QP safety-control framework. Section IV presents the simulation setup and performance evaluation, and Section V concludes the paper.
	
\section{Internal-Information-Free Performance-Equivalent Modeling of Closed-Loop Systems}
\label{sec2}
In this section, a performance-equivalent modeling framework for closed-loop systems is developed using only external port characteristics. The input-output closed-loop stability, generality, and non-uniqueness of feasible equivalent models for any closed-loop systems in DC microgrids are rigorously established, by combining the Small-Gain Theorem with frequency-domain analysis.
		
\subsection{Design Principle of the Performance-Equivalent Model}
	
Unlike conventional approaches that reconstruct subsystem dynamics from internal information or data, the proposed PEM-CLS defines a controllable closed-loop performance-equivalence boundary from port-level responses by reserving part of the available performance margin.

\begin{figure}[!t]
\centering
\includegraphics[width=0.4\textwidth]{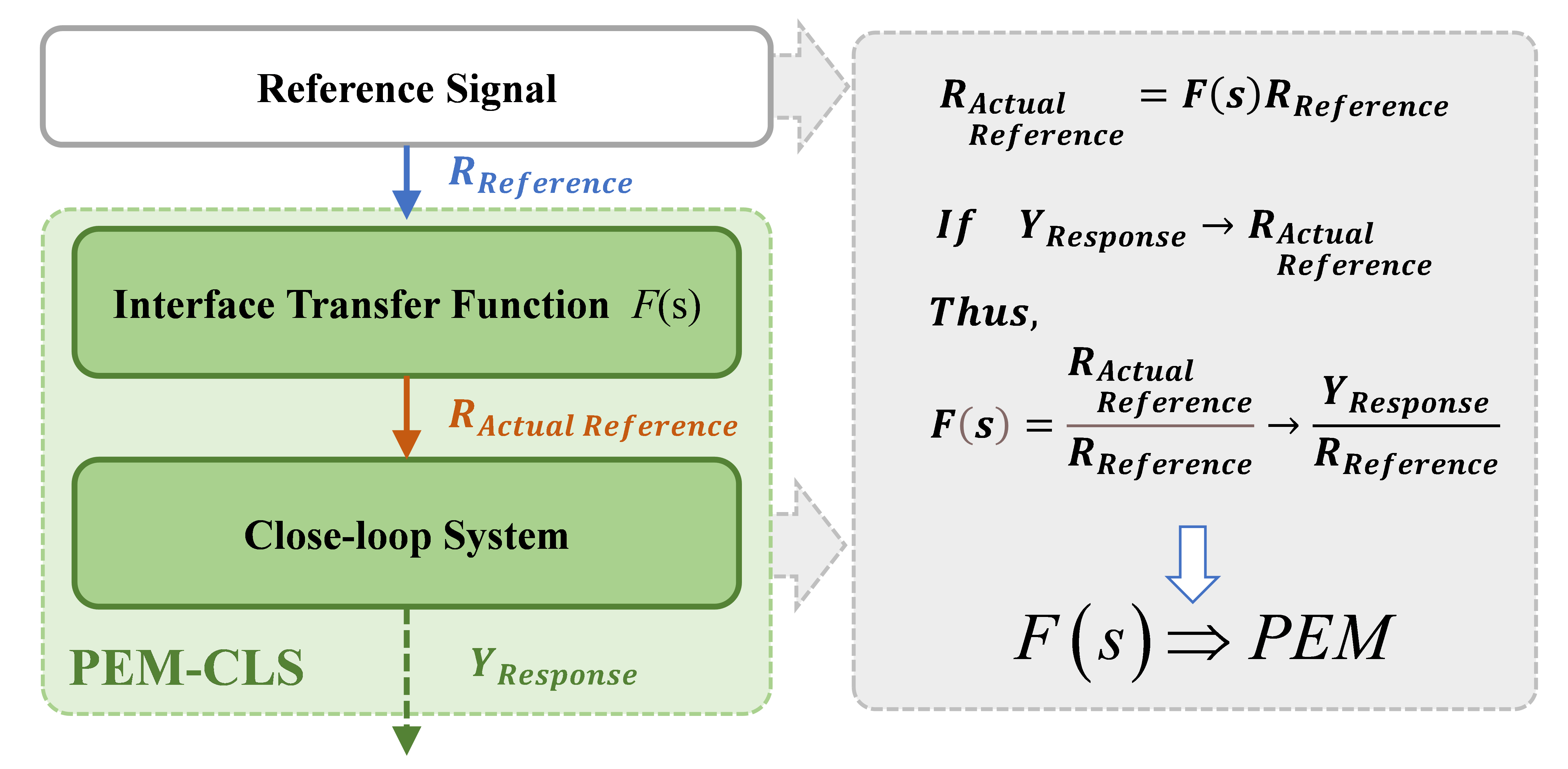}
\caption{Core concept of PEM-CLS for a closed-loop system.}
\label{Fig.1}
\end{figure}	
	
{\bf Core Concept}: As illustrated in Fig. \ref{Fig.1}, a performance-encapsulation interface function $F(s)$ is intentionally introduced between the upper-level scheduler and the local closed-loop control system. The interface limits the command magnitude of the upper-level dispatch applied to the subsystem to achieve dynamic matching, and only the processed command is used as the actual control command(power dispatch). The controlled system can therefore be equivalently represented by the combination of the $F(s)$  and the underlying subsystem.

When the subsystem can track the actual command with negligible error, or with an error acceptable to the upper-level scheduler, the designed interface function can be regarded as the external closed-loop PEM of the actual subsystem. Since the upper-level command, interface transfer function, actual command, and system response are all available in this architecture, the PEM modeling error can be directly quantified from the tracking response in terms of peak and average errors.

{\bf Design Rationale}: In practical digital control systems, upper-level commands are implemented as discrete steps; hence, the closed-loop command-tracking capability determines the admissible PEM equivalence boundary. Without explicit rate limits, a zero-to-full-load step within the regulation range can therefore serve as the benchmark signal for boundary identification. Within this boundary, PEMs with different orders and bandwidths can be constructed to balance EMS optimization requirements and computational complexity. Since both the benchmark signal and the $F(s)$ are known, PEM errors can be directly evaluated by comparing the closed-loop subsystem response with the corresponding interface-model response, yielding the peak and average equivalence errors.
\subsection{Proof of Existence and Non-Uniqueness of PEM}
	
	Various parameter perturbations (such as LCL filter drifts, line impedance variations, and PWM transport delays) and operational fluctuations jointly induce discrepancies between actual responses at the point of common coupling (PCC) and predictions from the nominal model. Under the per-unit framework, those physical mismatches inherently manifest in the frequency domain as relative proportional offsets with respect to the nominal transfer function. Therefore, guided by analysis methods of robust control theory, output multiplicative uncertainty is adopted to describe the modeling of these multi-source disturbances.
	
	Let $P(s)$ be the nominal system transfer function matrix. Incorporating the output multiplicative uncertainty, the actual physical object is formulated as
	\begin{equation}
		\label{equ1}
		{P_\Delta }(s) = \left[ {1 + {W_1}\left( s \right){\Delta _o}\left( s \right){W_2}\left( s \right)} \right]P(s)
	\end{equation}
	Where $W_1(s)$ and $W_2(s)$ serve as frequency-domain uncertainty weighting matrices, characterizing relative deviation bounds and inter-channel coupling across distinct frequency spectra. ${\Delta_{o}}\left( s \right)$ represents the normalized unknown dynamic perturbation, with ${\left\| {{\Delta_{o}}\left( s \right)} \right\|_\infty } \le 1$.
	
	\begin{assumption}
		\label{assum.1}
		The closed-loop controller $K$ (encompassing both linear and nonlinear controllers) guarantees stable operation of the nominal closed-loop system over the designated operating region. Within the designated operating range, local control action of $K$ around each operating point can be represented by a family of linear controllers $\left\{ {K_\rho(s):\rho  \in {{\rm P}_k}} \right\}$ that forms a well-performing closed-loop system with the nominal plant \cite{kaminer1995velocity}. Therefore, there exist a positive-definite matrix ${P_L}\left( \rho  \right)$ and a positive constant $\alpha > 0$ such that, for all $\rho  \in {{\rm P}_k}$ (operating point sets of the CLS), satisfy
		\begin{equation*}
			{{\bf{A}}_{{\bf{cl}}}}{\left( \rho  \right)^{\bf{T}}}{{\bf{P}}_{\bf{L}}}\left( \rho  \right) + {{\bf{P}}_{\bf{L}}}\left( \rho  \right){{\bf{A}}_{{\bf{cl}}}}\left( \rho  \right) \le  - \alpha {\bf{I}}
		\end{equation*}
	\end{assumption}
	
	\begin{assumption}
		\label{assum.2}
		The system closed-loop controller satisfies the small-gain robust stability condition; that is, there exists a constant $\mu< 1$ such that
		\begin{equation*}
			\mathop {\sup }\limits_{\rho  \in {\rm P}} {\left\| {{W_1}{W_2}{T_\rho }} \right\|_\infty } \le \mu  < 1
		\end{equation*}
	\end{assumption}
	
	\begin{remark}
	The above assumptions exhibit high applicability and generality for the microgrid lower-level control architecture. They only requires the subsystem to guarantee closed-loop asymptotic stability under design operating conditions, and its dynamic response can be equivalenced by a family of linear controllers. What's more, the performance of nonlinear control strategies around operating points can be bounded by a family of linear controllers.
	\end{remark}
	
	\begin{them}{}{robust-tracking}
		For each system stabilized by a feedback controller $K\left( s \right)$ within the target robust tracking frequency band ${\Omega _r}$, when the designed closed-loop performance equivalent function $F(s)$ satisfies
		\[
		\sup_{\rho \in \mathrm{P}}
		\sup_{\omega \in \Omega_r}
		\frac{
			\left|S_\rho(j\omega) F(j\omega)\right|
		}{
			1-
			\left|
			W_1(j\omega)
			\Delta_o(j\omega)
			W_2(j\omega)
			T_\rho(j\omega)
			\right|
		}
		\leq \mu \ll 1,
		\]
		\[{S_\rho } = \frac{1}{{1 + {L_\rho }}}; {T_\rho }\left( s \right) = \frac{{{L_\rho }}}{{1 + {L_\rho }}}; {L_\rho } = P(s){K_\rho }\left( s \right)\]
		Where, ${ L_\rho }$ represents the open-loop transfer function corresponding to an arbitrary fixed point; ${S_\rho }$ and ${T_\rho } $ are the nominal sensitivity function and complementary sensitivity function, respectively.\\ 
		
		The PEM of the system can be expressed as
		\[y(s)/r_c(s) = F\left( s \right)\]
	\end{them}
	
	{\bf{Proof:}} By Assumption \ref{assum.1}, $P_L$ is a positive definite matrix, such that for $\rho  \in {{\rm P}_k}$ and satisfy
	\begin{equation}
		\label{equ.3}
		\begin{array}{c}
			\dot V\left( x \right) = {x^T}\left( {{{\bf{A}}_{{\bf{cl}}}}{{\left( \rho  \right)}^{\bf{T}}}{{\bf{P}}_{\bf{L}}} + {{\bf{P}}_{\bf{L}}}{{\bf{A}}_{{\bf{cl}}}}\left( \rho  \right)} \right)x		\le  - \alpha {\left\| x \right\|^2}
		\end{array}
	\end{equation}
	where $V(x)=x^{T}P_Lx$. Thus, the nominal closed-loop system is uniformly stable over ${\rm P}_k$. Furthermore, Assumption \ref{assum.2} and ${\left\| 
		{{\Delta _o}} \right\|_\infty } \le 1$ give
	\begin{equation}
	\label{equ.4}
	{\left\| {{W_1}{W_2}{T_\rho }} \right\|_\infty }{\left\| {{\Delta _o}} \right\|_\infty } \le \mu  < 1
	\end{equation}

Therefore, by the small-gain theorem, the feedback interconnection with $\Delta_o$ is internally stable for every $\rho\in\mathcal{P}_k$. The corresponding tracking-error transfer function follows directly as
	\begin{equation}
		\frac{e}{r} = \frac{1}{{1 + {L_\rho }}}\frac{1}{{1 + {W_1}\left( s \right){\Delta _o}\left( s \right){W_2}\left( s \right){T_\rho }}},e = r - y 
	\end{equation}

Under closed-loop control, the upper bound of the maximum tracking error transfer function can be expressed as
	\begin{equation}
		\left| {{S_{\Delta ,\rho }}\left( {jw} \right)} \right| \le \frac{{\left| {{S_\rho }\left( {jw} \right)} \right|}}{{1 - \left| {{W_1}\left( {jw} \right){W_2}\left( {jw} \right){T_\rho }\left( {jw} \right)} \right|}},\forall w \in {\bf{R}}
	\end{equation}
	
	Because the original reference command is processed by $F(s)$ before being applied to the controlled subsystem as the actual reference input of the physical system, it follows that
	\begin{equation}
		r\left( s \right) = F\left( s \right){r_c}\left( s \right)
	\end{equation}
	
	Thus, the upper bound of the tracking error transfer function for $r_c$ can be
	\begin{equation}
		\begin{array}{l}
			\frac{{e\left( s \right)}}{{{r_{\rm{c}}}\left( s \right)}}  \le \mathop {\sup }\limits_{\rho  \in {\rm P}} \mathop {\sup }\limits_{w \in {\Omega _r}} \frac{{\left| {{S_\rho }\left( {jw} \right)F\left( {jw} \right)} \right|}}{{1 - \left| {{W_1}\left( {jw} \right){W_2}\left( {jw} \right){T_\rho }\left( {jw} \right)} \right|}}\le {\varepsilon _F}
		\end{array}
	\end{equation}
	
	When the ${\varepsilon_F}$ satisfies $0 < {\varepsilon_F} \ll 1 $, the PEM can be
	\begin{equation}
		{y\left( s \right)}/{{r_c}\left( s \right)} = F\left( s \right)
	\end{equation}
	
	{\bf {Proof ends.}}
	
	\begin{remark}
		The above proof is theoretically derived and proved for SISO systems, thus utilizing the complex modulus $|\cdot|$ in the frequency domain. When considering MIMO systems, the corresponding modulus can be simply converted to the maximum singular value, i.e., $|\cdot| \rightarrow \bar{\sigma}(\cdot)$.
	\end{remark}
	
	\begin{them}{}{robust-tracking}
		For any closed-loop control system, within the robust tracking bandwidth $\Omega_r$ of its feedback controller $K(s)$, there exist infinitely many sets of closed-loop performance-equivalent models $F_i(s)$ satisfying the design conditions in Theorem 1.
	\end{them}
	{\bf Proof:}
	Combining with Them~1, the condition for establishing the PEM is equivalent to $\left|S_{\Delta,\rho}F\right| \rightarrow 0$, which means	
	
    For $\omega \in \Omega_r$,
	\begin{equation}
		\label{equ.17}
		\begin{array}{c}
			\mathop {\sup }\limits_{w \in {\Omega _r}} \left| {F\left( {jw} \right)} \right| \le {M_F} < \infty ,\left| {{S_{\Delta ,\rho }}\left( {jw} \right)} \right| \ll 1\\
			\Rightarrow \left| {{S_{\Delta ,\rho }}F} \right| \to 0
		\end{array}  
	\end{equation}
	
	For $\omega \notin \Omega_r$,
	\begin{equation}
		\label{equ.18}
		\begin{array}{c}
			\mathop {\sup }\limits_{w \notin {\Omega _r}} \left| {F\left( {jw} \right)} \right| \ll 1,\left| {{S_{\Delta ,\rho }}\left( {jw} \right)} \right| \le {M_F} < \infty \\
			\Rightarrow \left| {{S_{\Delta ,\rho }}F} \right| \to 0
		\end{array}
	\end{equation}
For a given stabilized closed-loop system, there exists a family of models that are equivalent in terms of the prescribed closed-loop performance, denoted by$\mathcal{F}=\{F_i(s)\}$. For every $F_i(s)\in\mathcal{F}$ the conditions in Equ.\ref{equ.17}-\ref{equ.18} are satisfied, which yields
	\begin{equation}
		\frac{y(s)}{r_c(s)} = F_i(s).
	\end{equation}
	
	{\bf Proof ends.} 
	
	\section{PEM-Based HOCBF-CLF-QP Security Control for DC Microgrids}
	This section uses the performance-equivalent models developed in Section II to formulate an HOCBF–CLF–QP controller that coordinates subsystem power responses and imposes safety constraints during the execution of EMS dispatch commands. A representative topology of a medium-voltage DC (MVDC) microgrid is considered, as shown in Fig. 2. The DC bus acts as the central energy hub, to which individual subsystems are interfaced through power electronic converters, including rectifiers and inverters, for power exchange. The DC bus is also connected to the external utility grid through a grid-side inverter, enabling bidirectional power exchange between the microgrid and the utility grid. 
		
	The controller integrates three components: HOCBF-based safety constraints, CLF-based regulation constraints, and a QP for power-reference adjustment\cite{ames2016control}. First, HOCBFs are employed to enforce the safety constraints on the DC-bus voltage and supercapacitor SOC, thereby guaranteeing forward invariance of the safe sets. Second, CLFs are introduced to recover the voltage and SOC toward their desired values within the safe region. Third, the HOCBF and CLF constraints are coordinated through an online QP, where safety constraints are treated as hard constraints and performance objectives as soft constraints.
		\begin{figure}[!t]
		\centering
		\includegraphics[width=0.5\textwidth]{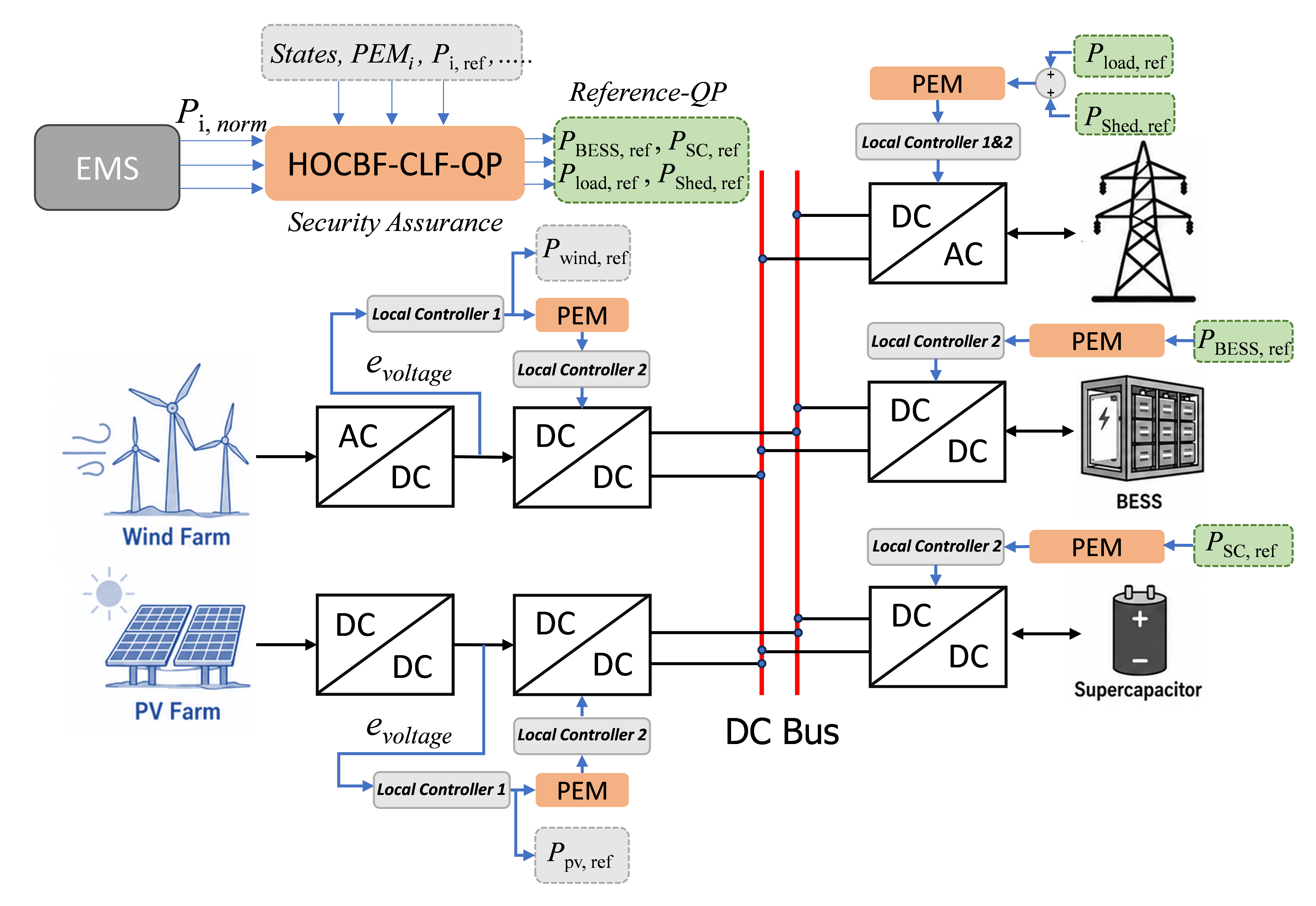}
		\caption{DC microgird system structure with PEM-based HOCBF-CLF-QP security control.}
		\label{Fig2}
	\end{figure}
	\subsection{HOCBF Formulation for Safety Constraints}
	To ensure that the DC-bus voltage remains within the prescribed safe operating range at all times, the voltage constraints are explicitly incorporated into the controller design. Specifically, HOCBFs are recursively constructed to transform the upper and lower voltage limits into safety inequalities that can be directly incorporated into the real-time control signal limitations, thereby guaranteeing forward invariance of the associated safe set\cite{xiao2021high}.
	
	Based on the modeling method presented in Section~~\ref{sec2}, the proposed PEM for the closed-loop system is applied to build equivalent models for the heterogeneous subsystems of the microgrid. In this section, the PEMs of these subsystems are uniformly represented by first-order lag dynamics with different time constants to accommodate their distinct dynamic response capabilities, i.e.,
	\begin{equation}
		\label{equ.20}
		{P_{i,ref}} = {\tau _i}\frac{\partial }{{\partial t}}\left( {{P_i}} \right) + {P_i}
	\end{equation}
	where $P_{i, ref}$ denotes the power target generated by the upper-level controller, and ${\tau _i}$ is the designed nominal response time constant characterizing the dynamic capability of subsystem $i$.

	{\bf{Robust HOCBF for DC Voltage:}}
	In the considered MVDC microgrid, the common DC bus is modeled by an equivalent lumped capacitance. Its energy dynamics are given by
	\begin{equation}
		\label{equ.22}
		\begin{array}{c}
			{E_{dc}} = \frac{1}{2}CV_{dc}^2\left( t \right)\\
			C{V_{dc}}{{\dot V}_{dc}} = \sum\limits_{i \in S} {{P_{i,actual}} - {P_{load}} + {P_{loss}}} 
		\end{array}
	\end{equation}
	where $C$ is the equivalent DC-bus capacitance, $S$ denotes the set of controllable subsystems, and $P_{loss}$ represents the aggregate power-loss term.
	
	The loss term $P_{loss}(V_{dc},I_{dc},t)$ accounts for line losses, e.g.,${P_{line}} = {{V_{dc}^2} \mathord{\left/{\vphantom {{V_{dc}^2} {{R_{eq}}}}} \right.\kern-\nulldelimiterspace} {{R_{eq}}}}$, and other unmodeled losses. Since $V_{dc}$ and $I_{dc}$ change continuously and practical power losses are bounded, the following assumption is introduced.
	
	\begin{assumption}
		$P_{loss}$ and $\dot P_{loss}$ are continuous and bounded \cite{ihammouchen2025nonlinear}, i.e., there exist constants $\mathchar'26\mkern-10mu\lambda _1>0$ and $\mathchar'26\mkern-10mu\lambda _2>0$, such that
		\begin{equation*}
			\left| {{P_{loss}}} \right| \le {\mathchar'26\mkern-10mu\lambda _1},\left| {{{\dot P}_{loss}}} \right| \le {\mathchar'26\mkern-10mu\lambda _2}
		\end{equation*}
	\end{assumption}
	
	To enforce the prescribed DC-bus voltage limits, define
	\begin{equation}
		\label{equ.23}
		\begin{array}{c}
			{X_{\min }} \le X \le {X_{\max }}\\
			{X_{\min }} \buildrel \Delta \over = V_{dc\_\min }^2,X = V_{dc}^2\left( t \right),{X_{\max }} \buildrel \Delta \over = V_{dc\_\max }^2
		\end{array}
	\end{equation}
	
	Based on Equ.\ref{equ.22}-\ref{equ.23}, the active-power balance at the common DC bus can be expressed as
	\begin{equation}
		\dot X = \frac{2}{C}\left( {\sum\limits_{i \in S} {{\sigma _i}{P_i}}  + {\sigma _{load}}{P_{load}} + {\sigma _{Shed}}{P_{Shed}} + {P_{loss}}} \right)
	\end{equation}
	Where ${\sigma _i} > 0$ and ${\sigma _i} < 0$ denote power injection into and extraction from the DC bus, respectively. The term $\sigma _{shed} P_{shed}$ represents emergency load shedding or power dumping under extreme operating conditions.
	
	According to the HOCBF framework \cite{xiao2021high}, a second-order HOCBF recursion is adopted to construct the auxiliary
	functions as
	\begin{equation}
		\varphi_{L0}=X-X_{\min}>0,
	\end{equation}
	\begin{equation}
		\varphi_{L1}=\frac{d}{dt}\varphi_{L0}+p_{L1}(X-X_{\min})^{q_{L1}}\geq 0\nonumber
	\end{equation}
	\begin{equation}
		\varphi_{L2}=\frac{d}{dt}\varphi_{L1}+p_{L2}\left(\varphi_{L1}\right)^{q_{L2}}\geq 0,
	\end{equation}
	where $p_{L1},p_{L2}>0$, $q_{L1}\geq1$, and $q_{L2}>0$.
    
	According to the forward-invariance property of HOCBFs, if the initial state lies within the safe set and $\varphi_{L2}\geq0$is enforced, the lower-bound safety condition $\varphi_{L0}=X-X_{\min}>0$ remains satisfied. Combining with the PEMs (Equ. \ref{equ.20}), the following second-order HOCBF condition holds
	\begin{equation}
		\begin{aligned}
			\varphi_{L2}
			={}&
			p_{L2}
			\left(
			p_{L1}e_L^{q_{L1}}
			+\frac{2\nu_1}{C}
			\right)^{q_{L2}}
			\\
			&+
			\frac{2p_{L1}q_{L1}}{C}
			e_L^{q_{L1}-1}\nu_1
			+\frac{2}{C}\dot{\nu}_1
			\geq0 .
		\end{aligned}
		\label{eq:hocbf_lower_original}
	\end{equation}
	where
	\begin{equation}
		e_L \triangleq X-X_{\min},
		\qquad
		\nu_1 \triangleq P_{\mathrm{loss}}+\nu_k,
	\end{equation}
	\begin{equation}
		\begin{aligned}
			\nu_k ={}&
			P_{\mathrm{load}}
			+P_{\mathrm{Shed}}
			+P_{\mathrm{pv}}
			+P_{\mathrm{wind}}
			+P_{\mathrm{bess}}
			+P_{\mathrm{SC}}
			+P_{\mathrm{gen}} .
		\end{aligned}
	\end{equation}
	
	Since the PV and wind power outputs are weather driven and cannot
	be directly manipulated by the safety controller,
	$P_{\mathrm{pv}}$, $P_{\mathrm{pv,ref}}$,
	$P_{\mathrm{wind}}$, and $P_{\mathrm{wind,ref}}$
	are treated as exogenous time-varying signals.
	The HOCBF condition therefore constrains only the power references
	of the controllable units.
	
	Define the control vector as
	\begin{equation}
		\mathbf{u}_c
		\triangleq
		\begin{bmatrix}
			P_{\mathrm{Shed,ref}} &
			P_{\mathrm{bess,ref}} &
			P_{\mathrm{SC,ref}} &
			P_{\mathrm{gen,ref}}
		\end{bmatrix}^{\mathrm T},
	\end{equation}
	with
	\begin{equation}
		\mathbf{a}_c
		\triangleq
		\frac{2}{C}
		\begin{bmatrix}
			\tau_{\mathrm{Shed}}^{-1} &
			\tau_{\mathrm{bess}}^{-1} &
			\tau_{\mathrm{SC}}^{-1} &
			\tau_{\mathrm{gen}}^{-1}
		\end{bmatrix}^{\mathrm T}.
	\end{equation}
	
	Furthermore, let
	\begin{equation}
		\begin{aligned}
			\eta_k \triangleq{}&
			\frac{P_{\mathrm{load,ref}}-P_{\mathrm{load}}}
			{\tau_{\mathrm{load}}}
			+
			\frac{P_{\mathrm{pv,ref}}-P_{\mathrm{pv}}}
			{\tau_{\mathrm{pv}}}
			\\
			&+
			\frac{P_{\mathrm{wind,ref}}-P_{\mathrm{wind}}}
			{\tau_{\mathrm{wind}}}
			-\frac{P_{\mathrm{bess}}}{\tau_{\mathrm{bess}}}
			-\frac{P_{\mathrm{SC}}}{\tau_{\mathrm{SC}}}
			-\frac{P_{\mathrm{Shed}}}{\tau_{\mathrm{Shed}}}
			-\frac{P_{\mathrm{gen}}}{\tau_{\mathrm{gen}}}.
		\end{aligned}
	\end{equation}
	
	The lower-bound HOCBF constraint can then be written as
	\begin{equation}
		\mathbf{a}_c^{\mathrm T}\mathbf{u}_c
		\geq
		H_L(P_{\mathrm{loss}},\dot P_{\mathrm{loss}}),
	\end{equation}
	where
	\begin{equation}
		\begin{aligned}
			H_L
			={}&
			-p_{L2}
			\left[
			p_{L1}e_L^{q_{L1}}
			+\frac{2(P_{\mathrm{loss}}+\nu_k)}{C}
			\right]^{q_{L2}}
			\\
			&-\frac{2p_{L1}q_{L1}}{C}
			e_L^{q_{L1}-1}
			(P_{\mathrm{loss}}+\nu_k)
			-\frac{2}{C}
			\left(
			\dot P_{\mathrm{loss}}+\eta_k
			\right).
		\end{aligned}
	\end{equation}
	
	By incorporating the boundedness conditions in Assumption~3, a robust real-time
	HOCBF constraint associated with the lower DC-bus voltage limit
	is formulated as
	\begin{equation}
		\mathbf{a}_c^{\mathrm T}\mathbf{u}_c
		\geq H_{\mathrm{dc-L}}
		\label{eq:robust_hocbf_lower}
	\end{equation}
	where
	\begin{equation}
		\begin{aligned}
			H_{\mathrm{dc-L}}
			\triangleq{}&
			-p_{L2}
			\left[
			p_{L1}e_L^{q_{L1}}
			+\frac{2(-\hat{\lambda}_1+\nu_k)}{C}
			\right]^{q_{L2}}
			\\
			&-\frac{2p_{L1}q_{L1}}{C}
			e_L^{q_{L1}-1}
			(-\hat{\lambda}_1+\nu_k)
			-\frac{2}{C}
			\left(-
			\hat{\lambda}_2+\eta_k
			\right).
		\end{aligned}
	\end{equation}
	
	Similarly, the robust HOCBF constraint associated with the upper
	DC-bus voltage limit can be expressed as
	\begin{equation}
		\mathbf{a}_c^{\mathrm T}\mathbf{u}_c
		\leq H_{\mathrm{dc-U}},
		\label{eq:robust_hocbf_upper}
	\end{equation}
	where
	\begin{equation}
		e_U \triangleq X_{\max}-X,
	\end{equation}
	\begin{equation}
		\begin{aligned}
			H_{\mathrm{dc-U}}
			\triangleq{}&
			p_{U2}
			\left[
			p_{U1}e_U^{q_{U1}}
			-\frac{2(\hat{\lambda}_1+\nu_k)}{C}
			\right]^{q_{U2}}
			\\
			&-\frac{2p_{U1}q_{U1}}{C}
			e_U^{q_{U1}-1}
			(\hat{\lambda}_1+\nu_k)
			-\frac{2}{C}
			\left(
			\hat{\lambda}_2+\eta_k
			\right).
		\end{aligned}
		\label{eq:Hdc_upper}
	\end{equation}
	
	Combining the lower- and upper-bound HOCBF constraints, the
	admissible dynamic safety-control bounds for the DC-bus voltage
	are given by
	\begin{equation}
		H_{\mathrm{dc-L}}
		\leq
		\mathbf{a}_c^{\mathrm T}\mathbf{u}_c
		\leq
		H_{\mathrm{dc-U}}
		\label{eq:dc_voltage_safety_bounds}
	\end{equation}
	
	{\bf{HOCBF for Suppercapacitor SOC:}}
	Owing to its limited energy capacity and high transient power capability, the supercapacitor is susceptible to overcharging and overdischarging during DC-bus power regulation. Therefore, its state of charge (SOC) is constrained within the prescribed safe operating range,
	\begin{equation}
		s_{\min}
		\leq s_{\mathrm{sc}}
		\leq s_{\max},s_{\mathrm{sc}}\triangleq SOC_{\mathrm{sc}}
	\end{equation}
	
	The rated energy capacity of the supercapacitor is given by
	\begin{equation}
		E_{\mathrm{sc}}
		=
		\frac{1}{2}C_{\mathrm{sc}}V_{\mathrm{norm}}^{2}.
	\end{equation}
	Defining $P_{\mathrm{sc}}>0$ as discharging power, the SOC dynamics
	are expressed as
	\begin{equation}
		\dot{s}_{\mathrm{sc}}
		=
		-\frac{P_{\mathrm{sc}}}{E_{\mathrm{sc}}}
		=
		-k_{\mathrm{sc}}P_{\mathrm{sc}},
		\qquad
		k_{\mathrm{sc}}
		\triangleq
		\frac{1}{E_{\mathrm{sc}}}.
	\end{equation}
	
	Since the power reference $P_{\mathrm{sc,ref}}$ enters the SOC dynamics through the PEM dynamics, the SOC safety function has relative degree two with respect to $P_{\mathrm{sc,ref}}$. For the lower SOC limit, define
	\begin{equation}
		h_{\mathrm{sc,L}}
		\triangleq
		s_{\mathrm{sc}}-s_{\min}.
	\end{equation}
	The corresponding second-order HOCBF is constructed as
	\begin{equation}
		\psi_{\mathrm{sc,L}}^{1}
		=
		-k_{\mathrm{sc}}P_{\mathrm{sc}}
		+
		p_{1,L}h_{\mathrm{sc,L}}^{q_{1,L}},
	\end{equation}
	\begin{equation}
		\psi_{\mathrm{sc,L}}^{2}
		=
		\dot{\psi}_{\mathrm{sc,L}}^{1}
		+
		p_{2,L}
		\left(
		\psi_{\mathrm{sc,L}}^{1}
		\right)^{q_{2,L}}
		\geq 0.
	\end{equation}
	\begin{equation}
		\tau_{\mathrm{sc}}\dot{P}_{\mathrm{sc}}
		+P_{\mathrm{sc}}
		=
		P_{\mathrm{sc,ref}},
	\end{equation}
	the HOCBF condition can be compactly expressed as
	\begin{equation}
		\frac{k_{\mathrm{sc}}}{\tau_{\mathrm{sc}}}
		P_{\mathrm{sc,ref}}
		\leq
		H_{\mathrm{sc-U}},
		\label{eq:soc_lower_hocbf}
	\end{equation}
	where
	\begin{equation}
		\begin{aligned}
			H_{\mathrm{sc-U}}
			\triangleq{}&
			\frac{k_{\mathrm{sc}}}{\tau_{\mathrm{sc}}}
			P_{\mathrm{sc}}
			-
			p_{1,L}q_{1,L}k_{\mathrm{sc}}P_{\mathrm{sc}}
			h_{\mathrm{sc,L}}^{q_{1,L}-1}
			\\
			&+
			p_{2,L}
			\left(
			-k_{\mathrm{sc}}P_{\mathrm{sc}}
			+
			p_{1,L}h_{\mathrm{sc,L}}^{q_{1,L}}
			\right)^{q_{2,L}} .
		\end{aligned}
		\label{eq:soc_upper_control_bound}
	\end{equation}
	
	This constraint limits the admissible discharging command as the SOC approaches its lower safety boundary, thereby preventing overdischarge.
	
	Similarly, the HOCBF constraint associated with the upper SOC limit yields the following lower bound on the supercapacitor power reference:
	\begin{equation}
		\frac{k_{\mathrm{sc}}}{\tau_{\mathrm{sc}}}
		P_{\mathrm{sc,ref}}
		\geq Q_L,
		\label{eq:soc_upper_hocbf}
	\end{equation}
	where
	\begin{equation}
		\begin{aligned}
			Q_L \triangleq{}&
			\frac{k_{\mathrm{sc}}}{\tau_{\mathrm{sc}}}
			P_{\mathrm{sc}}
			-
			p_{1,U}q_{1,U}k_{\mathrm{sc}}P_{\mathrm{sc}}
			(s_{\max}-s_{\mathrm{sc}})^{q_{1,U}-1}
			\\
			&-
			p_{2,U}
			\left[
			k_{\mathrm{sc}}P_{\mathrm{sc}}
			+
			p_{1,U}(s_{\max}-s_{\mathrm{sc}})^{q_{1,U}}
			\right]^{q_{2,U}} .
		\end{aligned}
		\label{eq:soc_lower_control_bound}
	\end{equation}
	
	Combining the lower- and upper-bound HOCBF constraints, the admissible safe control set for the supercapacitor is given by
	\begin{equation}
		\mathcal{U}_{\mathrm{safe}}
		=
		\left\{
		u_{\mathrm{sc}}\in\mathcal{U}
		\;\middle|\;
		Q_L \leq u_{\mathrm{sc}} \leq Q_U
		\right\},
		u_{\mathrm{sc}}
		\triangleq
		\frac{k_{\mathrm{sc}}}{\tau_{\mathrm{sc}}}
		P_{\mathrm{sc,ref}}.
	\end{equation}

	\subsection{CLF Formulation for Performance Constraints}
	To drive the DC-bus voltage and supercapacitor SOC toward their normal operating points while preserving QP feasibility, second-order CLF-based soft constraints are introduced.
	
	For $x_i\in\{V_{dc},s_{\mathrm{sc}}\}$, define the regulation error as
	\begin{equation}
		e_i=x_i-x_i^{\mathrm{ref}}.
		\label{eq:clf_error}
	\end{equation}
	
	Under the proposed PEM framework, the corresponding power reference enters the second derivative of $e_i$. Accordingly, the recovery constraint is formulated as
	\begin{equation}
		-\delta_{\mathrm{CLF-L},i}
		\leq
		\ddot e_i+c_{i,1}\dot e_i+c_{i,0}e_i
		\leq
		\delta_{\mathrm{CLF-U},i},
		\label{eq:clf_recovery}
	\end{equation}
	where $c_{i,1}>0$ and $c_{i,0}>0$ specify the desired convergence dynamics, while $\delta_{\mathrm{CLF-L},i},\delta_{\mathrm{CLF-U},i}\geq0$ are relaxation variables of QP.
	
	\textbf{CLF for DC bus voltage: }Substituting the DC-bus energy dynamics into the CLF recovery condition yields the direction-dependent voltage recovery constraints. Define
	\begin{equation}
		e_{\mathrm{dc}}
		\triangleq
		V_{\mathrm{dc}}-V_{\mathrm{dc}}^{\mathrm{ref}},
	\end{equation}
	and
	\begin{equation}
		\begin{aligned}
			u_{\mathrm{dc}}
			\triangleq
			\frac{2}{C}\Bigg(
			&\frac{P_{\mathrm{Shed,ref}}}{\tau_{\mathrm{Shed}}}
			+\frac{P_{\mathrm{bess,ref}}}{\tau_{\mathrm{bess}}}
			+\frac{P_{\mathrm{SC,ref}}}{\tau_{\mathrm{SC}}}
			+\frac{P_{\mathrm{gen,ref}}}{\tau_{\mathrm{gen}}}
			\Bigg).
		\end{aligned}
		\label{eq:udc_clf}
	\end{equation}
	
	For compactness, define
	\begin{equation}
		g_{\mathrm{dc}}^{+}
		\triangleq
		\frac{2}{C}
		\left(\hat{\lambda}_1+\nu_k\right),
		\qquad
		g_{\mathrm{dc}}^{-}
		\triangleq
		\frac{2}{C}
		\left(-\hat{\lambda}_1+\nu_k\right),
		\label{eq:gdc_pm}
	\end{equation}
	\begin{equation}
		d_{\mathrm{dc}}^{+}
		\triangleq
		\frac{2}{C}
		\left(\hat{\lambda}_2+\zeta_{\mathrm{dc}}\right),
		\qquad
		d_{\mathrm{dc}}^{-}
		\triangleq
		\frac{2}{C}
		\left(-\hat{\lambda}_2+\zeta_{\mathrm{dc}}\right),
		\label{eq:ddc_pm}
	\end{equation}
	where
	\begin{equation}
		\begin{aligned}
			\zeta_{\mathrm{dc}}
			\triangleq{}&
			\frac{P_{\mathrm{load,ref}}-P_{\mathrm{load}}}
			{\tau_{\mathrm{load}}}
			+\frac{P_{\mathrm{pv,ref}}-P_{\mathrm{pv}}}
			{\tau_{\mathrm{pv}}}
			\\
			&+\frac{P_{\mathrm{wind,ref}}-P_{\mathrm{wind}}}
			{\tau_{\mathrm{wind}}}
			-\frac{P_{\mathrm{bess}}}{\tau_{\mathrm{bess}}}
			-\frac{P_{\mathrm{SC}}}{\tau_{\mathrm{SC}}}
			-\frac{P_{\mathrm{Shed}}}{\tau_{\mathrm{Shed}}}
			-\frac{P_{\mathrm{gen}}}{\tau_{\mathrm{gen}}}.
		\end{aligned}
		\label{eq:zeta_dc}
	\end{equation}
	
	The two-sided CLF constraints are expressed as
	\begin{equation}
		\begin{cases}
			u_{\mathrm{dc}}
			-\delta_{\mathrm{CLF-U,dc}}
			\leq
			\ell_{\mathrm{CLF-U,dc}},
			& e_{\mathrm{dc}}\geq0,
			\\[1mm]
			-u_{\mathrm{dc}}
			-\delta_{\mathrm{CLF-L,dc}}
			\leq
			\ell_{\mathrm{CLF-L,dc}},
			& e_{\mathrm{dc}}<0,
		\end{cases}
		\label{eq:dc_clf}
	\end{equation}
	where
	\begin{align}
		\ell_{\mathrm{CLF-U,dc}}
		=&
		-c_{\mathrm{dc},1}g_{\mathrm{dc}}^{+}
		-c_{\mathrm{dc},0}e_{\mathrm{dc}}
		-d_{\mathrm{dc}}^{+},
		\\
		\ell_{\mathrm{CLF-L,dc}}
		=&
		c_{\mathrm{dc},1}g_{\mathrm{dc}}^{-}
		+c_{\mathrm{dc},0}e_{\mathrm{dc}}
		+d_{\mathrm{dc}}^{-}.
	\end{align}
	
	Similarly, substituting the supercapacitor SOC dynamics into the CLF recovery condition yields the two-sided SOC recovery constraints. Define
	\begin{equation}
		e_{\rm sc}
		\triangleq
		S_{\rm soc}-S_{\rm soc}^{\rm ref},
		\qquad
		u_{\rm sc}
		\triangleq
		\frac{k_s}{\tau_{\rm sc}}P_{\rm sc,ref}.
		\label{eq:soc_clf_def}
	\end{equation}
	
	The corresponding CLF constraints can be compactly expressed as
	\begin{equation}
		\begin{cases}
			-u_{\rm sc}
			-\delta_{\rm CLF-U,sc}
			\leq
			\ell_{\rm CLF-U,sc},
			& e_{\rm sc}\leq0,
			\\[1mm]
			u_{\rm sc}
			-\delta_{\rm CLF-L,sc}
			\leq
			\ell_{\rm CLF-L,sc},
			& e_{\rm sc}>0,
		\end{cases}
		\label{eq:soc_clf_constraints}
	\end{equation}
	where
	\begin{align}
		\ell_{\rm CLF-U,sc}
		\triangleq{}&
		-\frac{k_s}{\tau_{\rm sc}}P_{\rm sc}
		+k_s c_{{\rm s},1}P_{\rm sc}
		-c_{{\rm s},0}e_{\rm sc},
		\label{eq:soc_clf_upper}
		\\
		\ell_{\rm CLF-L,sc}
		\triangleq{}&
		\frac{k_s}{\tau_{\rm sc}}P_{\rm sc}
		-k_s c_{{\rm s},1}P_{\rm sc}
		+c_{{\rm s},0}e_{\rm sc}.
		\label{eq:soc_clf_lower}
	\end{align}
	
	\subsection{HOCBF-CLF-Based QP Formulation}
	
	The HOCBF safety constraints and CLF recovery constraints are
	integrated into a unified QP for online control allocation.	The HOCBF constraints are imposed as hard constraints, whereas	the CLF constraints are softened by nonnegative relaxation	variables to preserve feasibility.
	
	Define the controllable power-reference vector as
	\begin{equation}
		\mathbf{u}
		\triangleq
		\begin{bmatrix}
			P_{\mathrm{bess,ref}} &
			P_{\mathrm{diesel,ref}} &
			P_{\mathrm{sc,ref}} &
			P_{\mathrm{Shed,ref}}
		\end{bmatrix}^{\mathrm T},
		\label{eq:qp_u}
	\end{equation}
	where $\mathbf{u}_{\mathrm{norm}}$ denotes the nominal power command
	provided by EMS. The CLF relaxation vector and the QP decision vector are defined as
	\begin{equation}
		\boldsymbol{\delta}
		\triangleq
		\begin{bmatrix}
			\delta_{\mathrm{CLF-U,dc}} &
			\delta_{\mathrm{CLF-L,dc}} &
			\delta_{\mathrm{CLF-U,sc}} &
			\delta_{\mathrm{CLF-L,sc}}
		\end{bmatrix}^{\mathrm T}
		\label{eq:qp_delta}
	\end{equation}
	\begin{equation}
		\boldsymbol{\zeta}
		\triangleq
		\begin{bmatrix}
			\mathbf{u}^{\mathrm T} &
			\boldsymbol{\delta}^{\mathrm T}
		\end{bmatrix}^{\mathrm T}.
		\label{eq:qp_zeta}
	\end{equation}
	
	Thus, the proposed HOCBF-CLF-QP is formulated as
	\begin{equation}
		\begin{aligned}
			\min_{\boldsymbol{\zeta}}\quad
			&
			\frac{1}{2}
			(\mathbf{u}-\mathbf{u}_{\mathrm{norm}})^{\mathrm T}
			H_u
			(\mathbf{u}-\mathbf{u}_{\mathrm{norm}})
			\\
			&+
			\rho_{\mathrm{Shed}}P_{\mathrm{Shed,ref}}^{2}
			+
			\boldsymbol{\delta}^{\mathrm T}
			R_{\delta}\boldsymbol{\delta}
			\\
			\mathrm{s.t.}\quad
			&
			A_{\mathrm{sc}}\boldsymbol{\zeta}
			\leq
			b_{\mathrm{sc}},
			\\
			&
			\mathbf{u}_{\min}
			\leq
			\mathbf{u}
			\leq
			\mathbf{u}_{\max},
			\\
			&
			|\Delta P_{\mathrm{diesel,ref}}|
			\leq
			R_{\mathrm{diesel}}^{\max},
			\qquad
			\boldsymbol{\delta}\geq\mathbf{0},
		\end{aligned}
		\label{eq:clf_hocbf_qp}
	\end{equation}
	where $H_u\succ0$ penalizes deviations from the nominal command,
	$R_{\delta}\succ0$ penalizes CLF relaxation, and
	$\rho_{\mathrm{Shed}}>0$ suppresses unnecessary power shedding.
	Their weighting factors can be selected to assign different
	priorities to voltage and SOC recovery.
	
	For a compact representation, define
	\begin{equation}
		\begin{aligned}
			\mathbf{a}_{\mathrm{dc}}^{\mathrm T}
			&\triangleq
			\frac{2}{C}
			\begin{bmatrix}
				\tau_{\mathrm{bess}}^{-1} &
				\tau_{\mathrm{diesel}}^{-1} &
				\tau_{\mathrm{sc}}^{-1} &
				\tau_{\mathrm{Shed}}^{-1}
			\end{bmatrix},
			\\
			\mathbf{a}_{\mathrm{sc}}^{\mathrm T}
			&\triangleq
			\begin{bmatrix}
				0 & 0 & k_s/\tau_{\mathrm{sc}} & 0
			\end{bmatrix}.
		\end{aligned}
		\label{eq:qp_coefficients}
	\end{equation}
Thus, the overall safety and performance constraints can therefore be assembled compactly as
	\begin{equation}
		A_{\mathrm{sc}}
		=
		\begin{bmatrix}
			A_{\mathrm H} & \mathbf{0}_{4\times4}\\
			A_{\mathrm C} & -I_4
		\end{bmatrix},
		\qquad
		b_{\mathrm{sc}}
		=
		\begin{bmatrix}
			\mathbf{b}_{\mathrm H}\\
			\mathbf{b}_{\mathrm C}
		\end{bmatrix}
		\label{eq:qp_Asc_bsc}
	\end{equation}
	\begin{equation}
		A_{\mathrm H}
		\triangleq
		\begin{bmatrix}
			\mathbf{a}_{\mathrm{dc}}^{\mathrm T}\\
			-\mathbf{a}_{\mathrm{dc}}^{\mathrm T}\\
			\mathbf{a}_{\mathrm{sc}}^{\mathrm T}\\
			-\mathbf{a}_{\mathrm{sc}}^{\mathrm T}
		\end{bmatrix},
		\qquad
		\mathbf{b}_{\mathrm H}
		\triangleq
		\begin{bmatrix}
			H_{\mathrm{dc-U}}\\
			-H_{\mathrm{dc-L}}\\
			Q_U\\
			-Q_L
		\end{bmatrix}.
		\label{eq:qp_hocbf}
	\end{equation}
	
	\begin{equation}
		A_{\mathrm C}
		\triangleq
		\begin{bmatrix}
			\mathbf{a}_{\mathrm{dc}}^{\mathrm T}\\
			-\mathbf{a}_{\mathrm{dc}}^{\mathrm T}\\
			-\mathbf{a}_{\mathrm{sc}}^{\mathrm T}\\
			\mathbf{a}_{\mathrm{sc}}^{\mathrm T}
		\end{bmatrix},
		\qquad
		\mathbf{b}_{\mathrm C}
		\triangleq
		\begin{bmatrix}
			\ell_{\mathrm{CLF-U,dc}}\\
			\ell_{\mathrm{CLF-L,dc}}\\
			\ell_{\mathrm{CLF-U,sc}}\\
			\ell_{\mathrm{CLF-L,sc}}
		\end{bmatrix}.
		\label{eq:qp_clf}
	\end{equation}
	
	\section{Case Study}
	
    To evaluate the effectiveness of the proposed safety execution layer under high-risk operating conditions, this study adopts a set of extreme stress-test scenarios. Compared with normal operating conditions, these scenarios are more likely to expose conflicts between EMS dispatch commands and the actual dynamic safety of the physical system, and therefore provide a more direct assessment of when and how the safety layer must intervene. The tests focus on whether the proposed method can promptly correct unsafe dispatch commands and maintain dynamic safety under these challenging conditions.
	\subsection{Simulation Configuration and Test Setup}
	\subsubsection{System and PEM Configuration}
	
	Following the PEM-CLS framework in Section~II, each subsystem is represented by a first-order performance-equivalent model $F(s)$ with an individual time constant. The main operating limits and PEM parameters are summarized in Table~\ref{tab:PEM_parameters}, and the overall MVDC microgrid and control architecture are shown in Fig.~\ref{Fig2}. 
	
	\begin{table}[t]
		\caption{Operating Limits and PEM Parameters}
		\label{tab:PEM_parameters}
		\centering
		\scriptsize
		\setlength{\tabcolsep}{2.5pt}
		\renewcommand{\arraystretch}{1.05}
		\begin{tabularx}{\columnwidth}{@{}lXc@{}}
			\hline
			System & Operating limits & $\tau_i$ \\
			\hline
			DC bus
			& $2.85\!-\!3.15$ kV ($\pm5\%$);
			$2.70\!-\!3.30$ kV ($\pm10\%$)
			& -- \\
			
			PV
			& $0\leq P_{\rm pv}\leq1$ MW
			& $5$ ms \\
			
			Wind
			& $0\leq P_{\rm wind}\leq1$ MW
			& $5$ ms \\
			
			BESS
			& $-0.5\leq P_{\rm bess}\leq0.5$ MW
			& $1$ s \\
			
			SC
			& $-1\leq P_{\rm sc}\leq1$ MW,
			$0.10\leq S_{\rm sc}\leq0.95$
			& $0.2$ s \\
			
			Generator
			& $0.05\leq P_{\rm gen}\leq0.5$ MW,
			$|\Delta P_{\rm gen}|\leq0.05$ MW
			& $5$ s \\
			
			Load/Shed
			& $-2\leq P_{\rm load},P_{\rm Shed}\leq2$ MW
			& $0.2$ s \\
			\hline
		\end{tabularx}
	\end{table}
	
\subsubsection{QP Configuration}

The QP parameters follow the safety-first design principle, with strong penalties on unnecessary load shedding and deviations from nominal EMS commands. Since the PEM interface function $F(s)$ decouples upper-level power dispatch from subsystem-level power-response implementation, the key system and controller parameters required to reproduce the proposed method are reported in Table~\ref{tab:control_parameters}.
	
	\begin{table}[t]
		\caption{Key System and Controller Parameters}
		\label{tab:control_parameters}
		\centering
		\scriptsize
		\setlength{\tabcolsep}{3pt}
		\renewcommand{\arraystretch}{1.08}
		\begin{tabularx}{\columnwidth}{@{}lX@{}}
			\hline
			Category & Parameters \\
			\hline
			
			System
			&
			$V_{\rm dc,nom}=3$ kV,
			$C_{\rm Bus-dc}=60$ mF;
			$V_{\rm sub,nom}=1.2$ kV,
			$C_{\rm SubBus-dc}=50$ mF
			\\
			
			HOCBF--DC
			&
			$p_{L1}=p_{U1}=3$,
			$p_{L2}=p_{U2}=10$;
			$q_{L1}=q_{U1}=1$,
			$q_{L2}=q_{U2}=1/3$;
			$\hat{\lambda}_{1}=\hat{\lambda}_{2}=0.05$
			\\
			
			HOCBF--SC
			&
			$p_{1,L}=p_{1,U}=2$;
			$q_{1,L}=q_{2,L}=p_{2,L}=p_{2,U}=q_{1,U}=q_{2,U}=1$;
			$k_{\rm sc}=0.4193$,
			$SOC_{\min}=0.10$,
			$SOC_{\max}=0.95$
			\\
			
			CLF--DC
			&
			$c_{{\rm dc},1}=10$,
			$c_{{\rm dc},0}=50$,
			$V_{\rm dc}^{\rm ref}=3$ kV
			\\
			
			CLF--SC
			&
			$c_{{\rm sc},1}=2$,
			$c_{{\rm sc},0}=4$,
			$S_{\rm sc}^{\rm ref}=0.5$
			\\
			
			QP weights
			&
			$w_{\rm bess}=2$,
			$w_{\rm gen}=3.33$,
                $w_{\rm sc}=2$,
			$w_{\rm Shed}=10^{3}$;
			$\delta_{\mathrm{CLF-U,dc}}=10^{6}$,
                $\delta_{\mathrm{CLF-L,dc}}=10^{6}$,
			$\delta_{\mathrm{CLF-U,sc}}=10^{3}$,
                $\delta_{\mathrm{CLF-L,sc}}=10^{3}$
			\\
			
			QP update
			&
			$T_{\rm QP}=0.1$ s
			\\
			\hline
		\end{tabularx}
	\end{table}

\begin{figure*}[!t]
	\centering
	\includegraphics[width=0.8\textwidth]{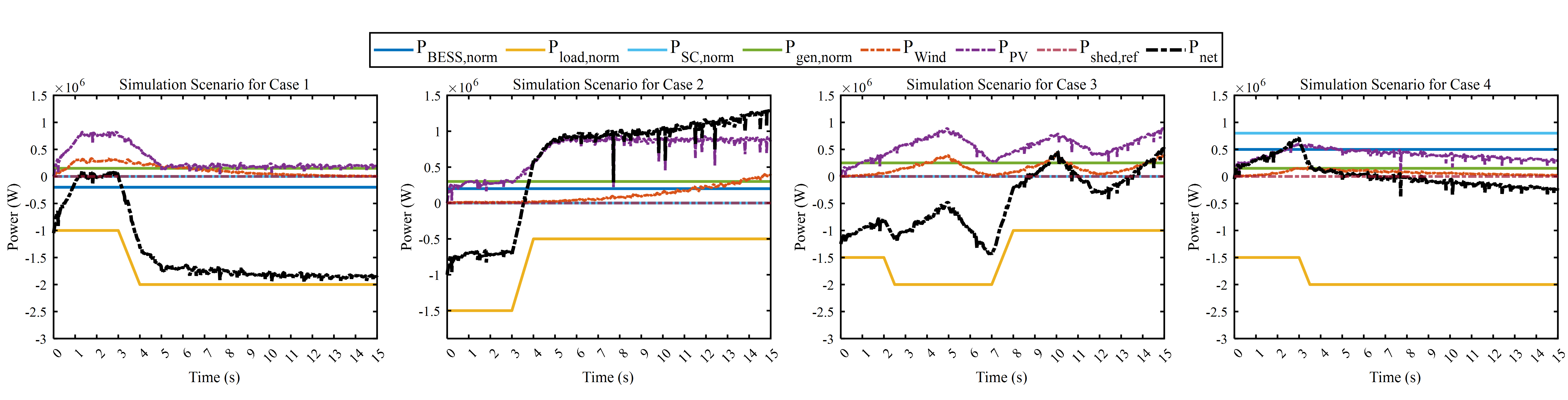}
	\caption{Stress-test scenarios and nominal power dispatch settings by EMS. {\bf{Note:}} ${\bf{P_{net}}}$ represents the net variable power of the DC bus.}
	\label{Fig3}
\end{figure*}

\begin{figure*}[!t]
	\centering
	\includegraphics[width=0.8\textwidth]{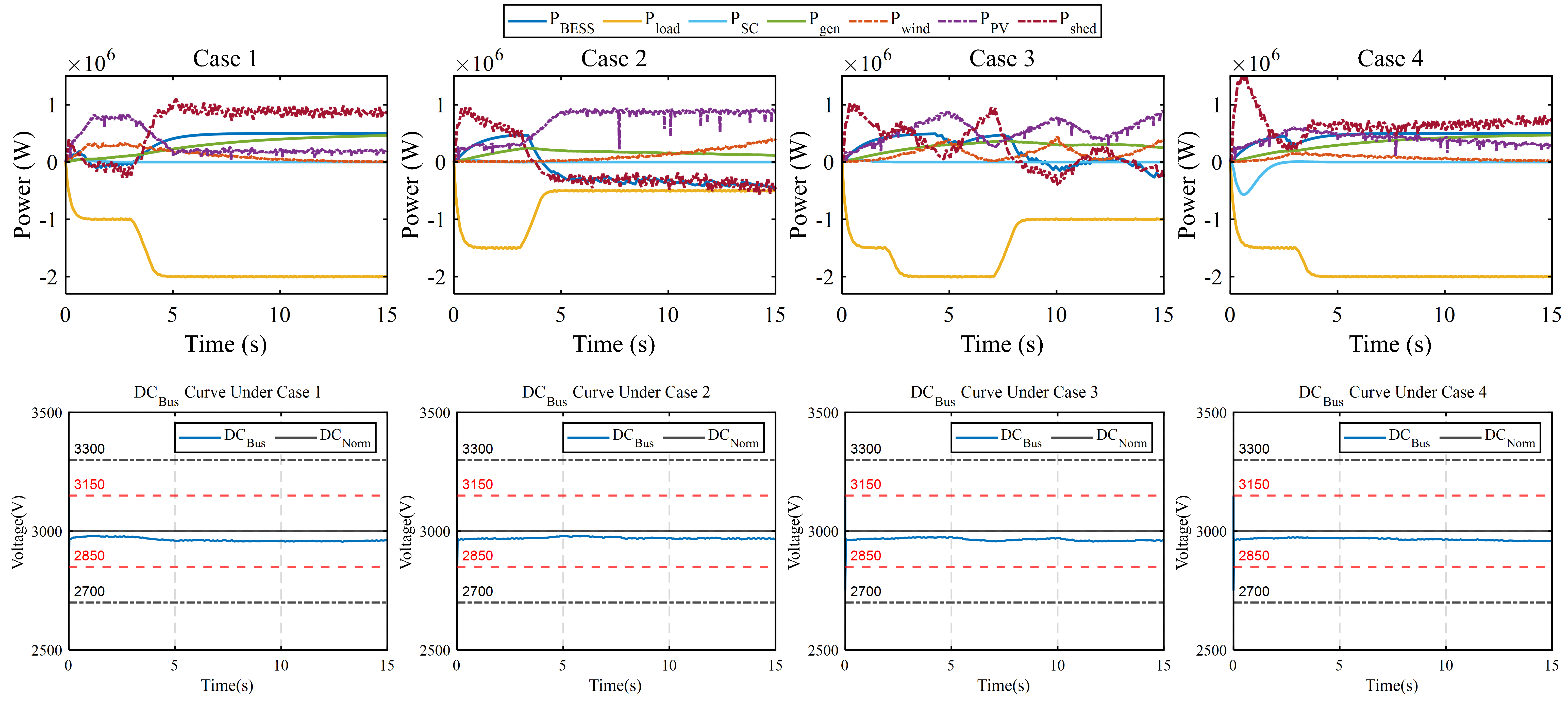}
	\caption{Actual power response of the DC microgrid subsystem controlled by HOCBF-CLF-QP and the voltage curve of the DC Bus.}
	\label{Fig4}
\end{figure*}

\begin{figure*}[!t]
	\centering
	\includegraphics[width=0.85\textwidth]{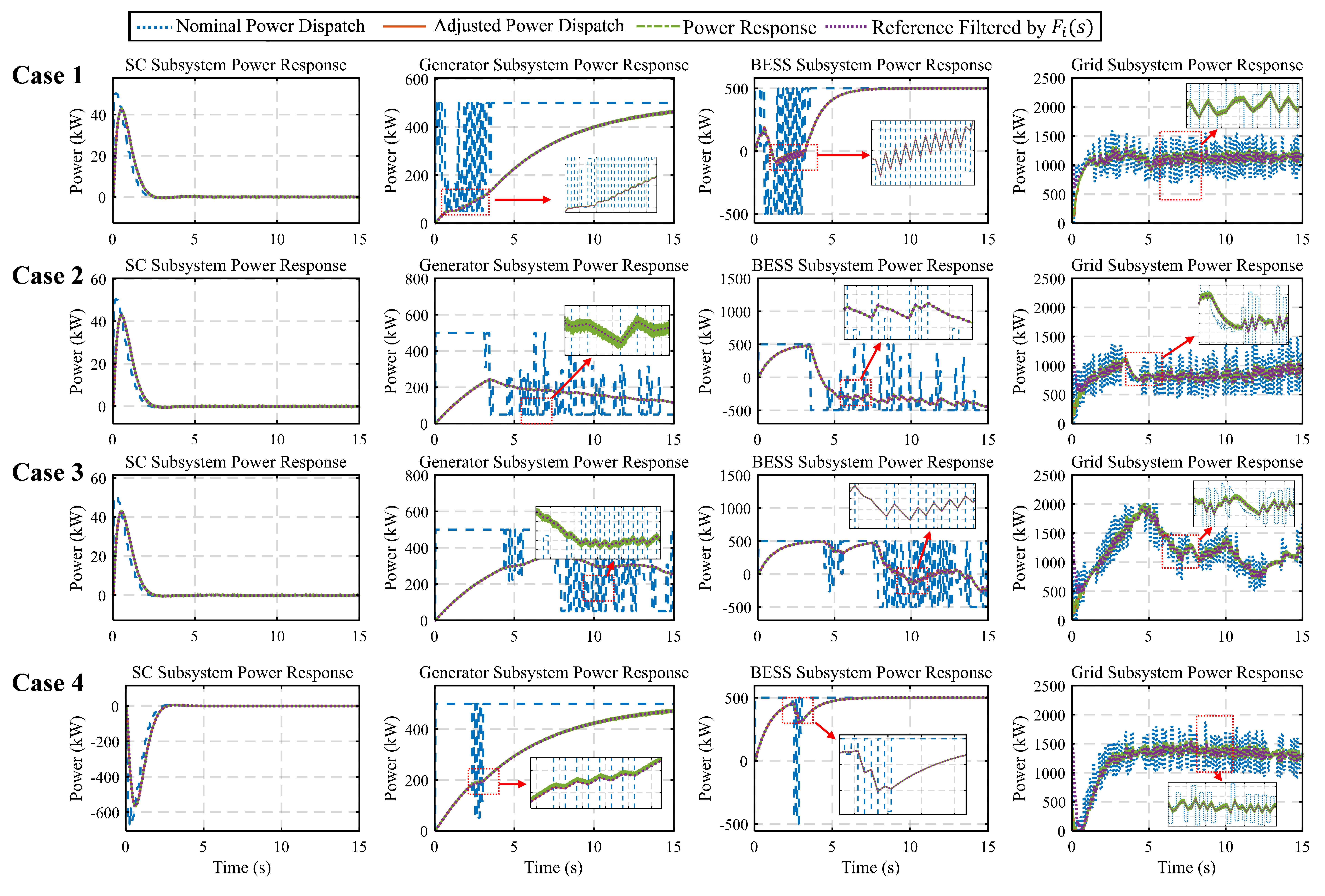}
	\caption{Actual power response of controlled subsystems of the DC microgrid.}
	\label{Fig5}
\end{figure*}	
	
\subsubsection{Stress-Test Scenarios}

Four stress-test scenarios evaluate the proposed method under severe disturbances and mismatches between nominal EMS commands and safety-feasible power commands. Each 15-s scenario includes high-frequency random renewable-power fluctuations of up to 100~kW, with the corresponding operating profiles shown in Fig.~\ref{Fig3}.

	\begin{figure*}[!t]
	\centering
	\includegraphics[width=0.85\textwidth]{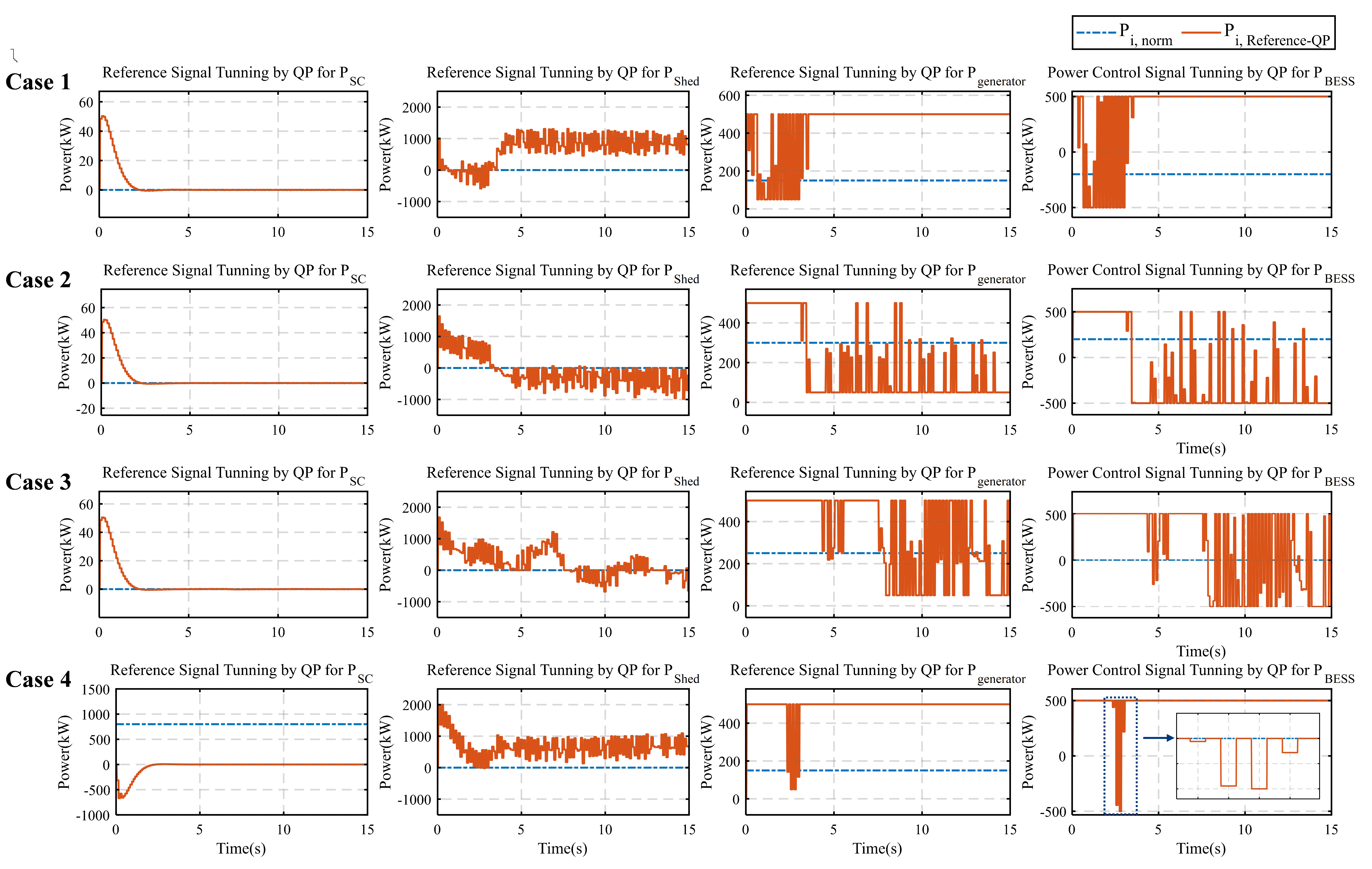}
	\caption{Power norm control signal tuning by QP for controlled subsystems in the DC microgrid system, $i=$SC, Shed, Generator, and BESS.}
	\label{Fig6}
\end{figure*}

\begin{figure*}[t]
	\centering
	
	\begin{minipage}[t]{0.68\textwidth}
		\centering
		\includegraphics[width=\linewidth]{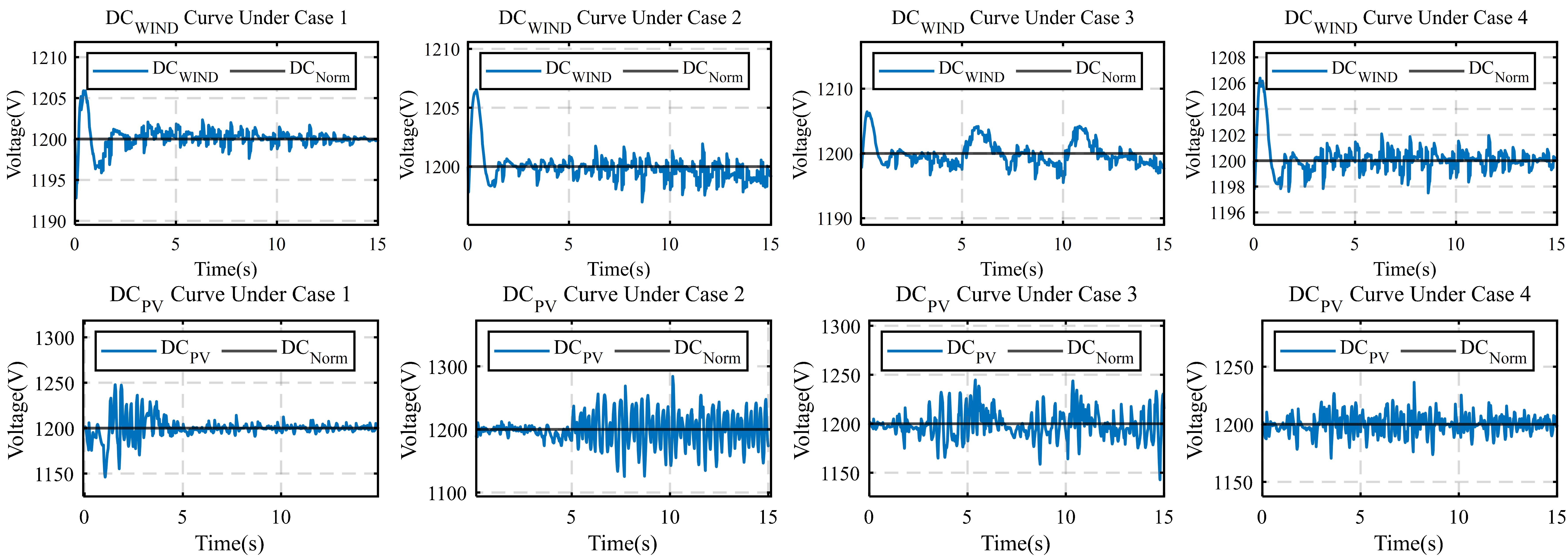}
		\caption{DC voltage curves of PV and Wind.}
		\label{Fig7}
	\end{minipage}
	\hfill
	\begin{minipage}[t]{0.29\textwidth}
		\centering
		\includegraphics[width=\linewidth]{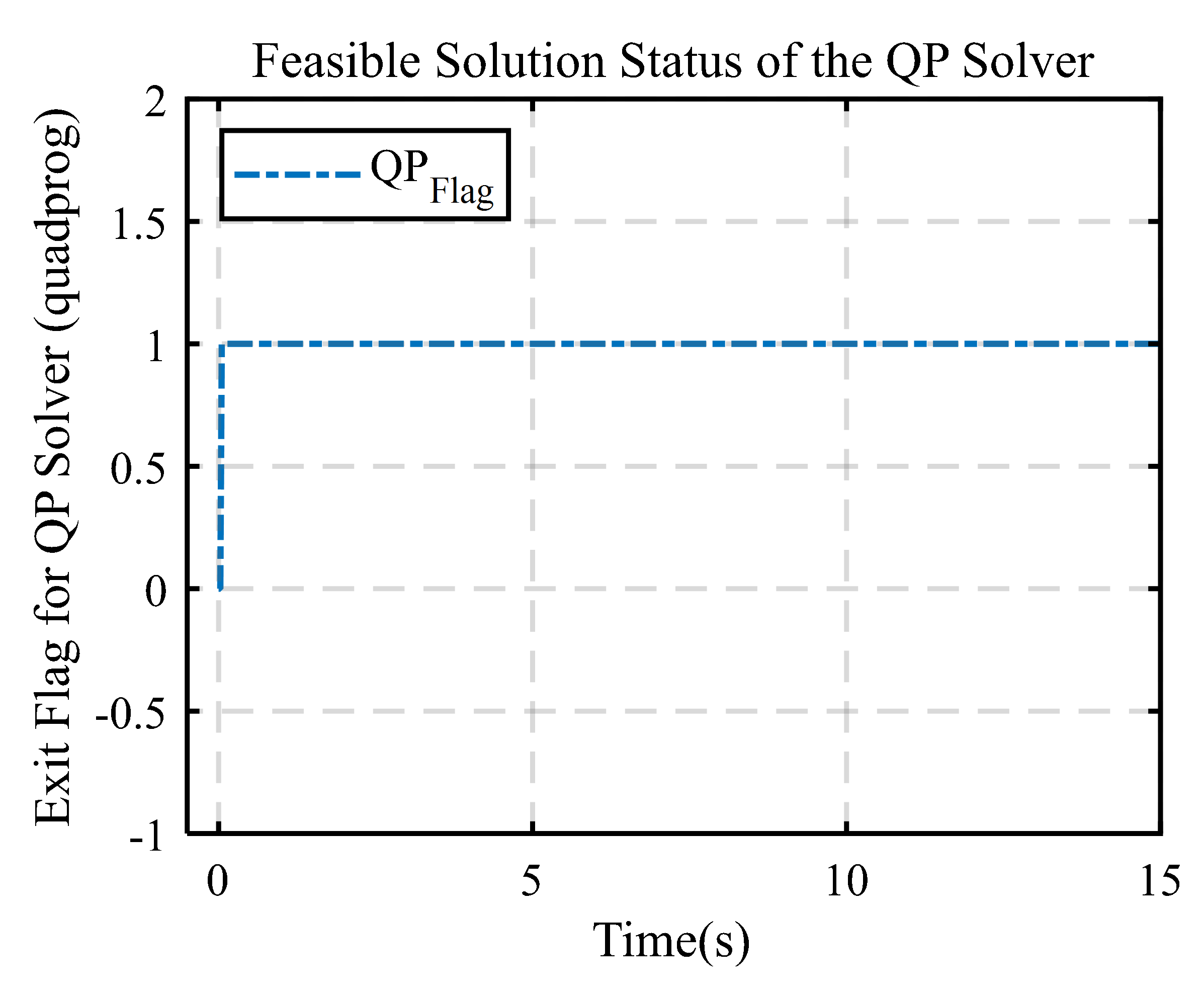}
		\caption{QP feasibility curve. The QP activates for the first time at t=0.05s; Flag=1 indicates that a feasible QP solution exists.}
		\label{Fig8}
	\end{minipage}
	
\end{figure*}

	\begin{figure*}[!t]
		\centering
		\includegraphics[width=1\textwidth]{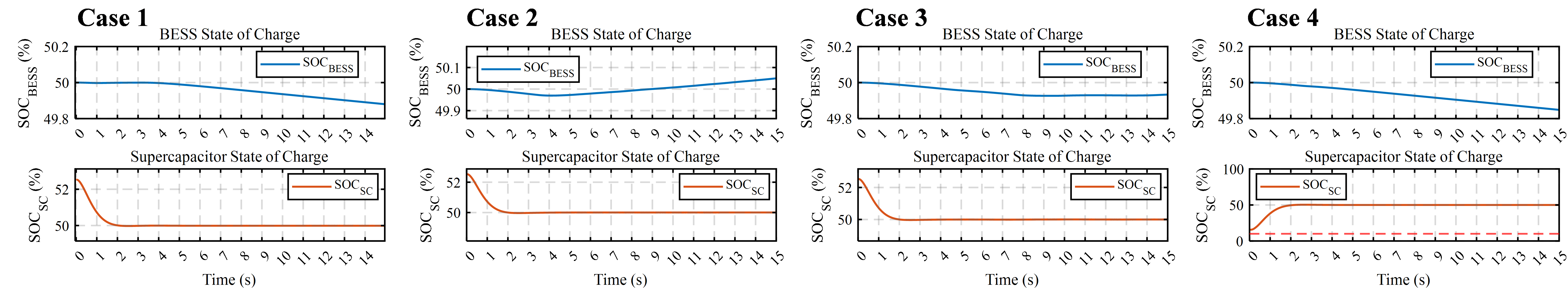}
		\caption{SOC curves of BESS and Suppercapacitor.}
		\label{Fig9}
	\end{figure*}

	\subsection{Simulation Results and Performance Analysis}
	
The proposed control ensures DC-bus voltage safety, corrects unsafe EMS commands, and enables dynamic tracking of the \(F_i(s)\)-adjusted power dispatch commands, while maintaining QP feasibility and regulating SOC. Without intervention, the DC-bus voltage violates its safety bound within 0.2~s in all cases; with control, its deviation remains within approximately \(\pm2\%\), with feasible QP solutions (Fig.~\ref{Fig8}) and close command tracking (Fig.~\ref{Fig5}).

\subsubsection{Case 1: Power Deficit}
Case~1 considers a severe generation deficit. The controller modifies the nominal power dispatch and coordinates the BESS, generator, and power shedding to restore power balance. The DC-bus voltage remains within the prescribed safety region, with a maximum deviation below 2\% (Fig.~\ref{Fig4}), while the PV- and wind-side DC-bus voltages remain within $\pm5\%$ (Fig.~\ref{Fig7}). The BESS, SC, generator, and grid interface accurately track their $F(s)$-adjusted power dispatch. After providing transient support, the SC SOC is driven back toward its 50\% reference (Fig.~\ref{Fig9}), thereby avoiding sustained SC discharge and preserving its energy margin for subsequent fast transients.

\subsubsection{Case 2: Power Deficit to Power Surplus}
Case~2 evaluates a rapid transition from a severe deficit to a power surplus caused by increased renewable generation and reduced load. Using the PEM-based HOCBF–CLF–QP controller, the controllable power references are adjusted to compensate for the power surplus. The resulting DC-bus voltage deviation is limited to approximately 1.8\% (Fig.~\ref{Fig4}), and the PV- and wind-side voltages remain within $\pm8\%$ (Fig.~\ref{Fig7}). The BESS correspondingly switches to charging operation, while all controllable subsystems continue to track their adjusted power dispatch with negligible tracking error.

\subsubsection{Case 3: Severe Power Fluctuation}
Case~3 introduces frequent bidirectional renewable-power variations exceeding 50\% of the subsystem rated power within 2~s, together with high-frequency random disturbances (Fig.~\ref{Fig3}). Despite these rapid variations, the DC-bus voltage is confined within approximately $\pm2\%$ of its nominal value (Fig.~\ref{Fig4}). The subsystem power responses, SC SOC, and other safety-related states remain well regulated, confirming the disturbance-rejection capability of the proposed coordination framework under highly fluctuating operating conditions.

\subsubsection{Case 4: Limited Resources and Unsafe Commands}
Case~4 evaluates the controller under limited energy reserves and an unsafe EMS command. Starting from a low-SOC condition (Fig.~\ref{Fig9}), direct execution of the nominal EMS command would continuously discharge the SC and eventually violate its SOC constraint. The HOCBF-CLF-QP controller instead corrects the unsafe command and reallocates power among the available controllable units (Figs.~\ref{Fig5} and \ref{Fig6}). During the initial surplus interval, the CLF objective drives the SC SOC toward its 50\% reference, after which the remaining power imbalance is accommodated by the other subsystems. Consequently, both the SC SOC and DC-bus voltage remain within their prescribed safety regions.

\section{Conclusion}
This paper proposed a PEM-CLS-based HOCBF-CLF-QP to coordinate subsystem power responses and adjust EMS dispatch commands for dynamically safe execution in DC microgrids. PEM-CLS captures the closed-loop dynamics required for safety-critical control solely from port-level responses, without requiring internal topology, controller parameters, or detailed physical models. Based on the resulting PEMs, the proposed strategy coordinates heterogeneous power responses and corrects unsafe EMS commands during command transitions. Under quantified modeling-error bounds and actuator constraints, the HOCBF constraints guarantee forward invariance of the DC-bus voltage safe set, providing a theoretical basis for dynamically safe dispatch execution. Simulations under severe disturbances and limited regulation capability maintained DC-bus voltage deviations within ±2\%. Overall, the proposed framework introduces a dynamic safety execution layer without altering the original EMS dispatch objective, extending steady-state feasible dispatch to verifiably safe dynamic execution in heterogeneous DC microgrids.

\bibliographystyle{IEEEtran}
\bibliography{ref}

\end{document}